\documentclass[sigconf, nonacm]{acmart}
\usepackage{cleveref}

\newcommand\vldbdoi{10.14778/3827998.3828086}
\newcommand\vldbpages{4642 - 4645}
\newcommand\vldbvolume{19}
\newcommand\vldbissue{12}
\newcommand\vldbyear{2026}
\newcommand\vldbauthors{\authors}
\newcommand\vldbtitle{\shorttitle} 
\newcommand\vldbavailabilityurl{https://github.com/mitdbg/carnot}
\newcommand\vldbpagestyle{empty}
\newcommand{\system}{{\sc Carnot}}

\begin{document}
\title{Carnot: Interpretable, Interactive, and Optimized Execution of Deep Research Queries}

\author{Matthew Russo}
\orcid{0009-0005-9685-3976}
\affiliation{%
  \institution{MIT}
}
\email{mdrusso@csail.mit.edu}

\author{Yash Agarwal}
\affiliation{%
  \institution{MIT}
}
\email{yashaga@mit.edu}

\author{Tianyu Li}
\affiliation{%
  \institution{MIT}
}
\email{litianyu@mit.edu}

\author{Zhuohan Gu}
\affiliation{%
  \institution{MIT}
}
\email{zgu15@mit.edu}

\author{Michael Cafarella}
\affiliation{%
  \institution{MIT}
}
\email{michjc@csail.mit.edu}

\author{Omar Khattab}
\affiliation{%
  \institution{MIT}
}
\email{okhattab@mit.edu}

\author{Tim Kraska}
\affiliation{%
  \institution{MIT}
}
\email{kraska@mit.edu}

\author{Samuel Madden}
\affiliation{%
  \institution{MIT}
}
\email{madden@csail.mit.edu}


\begin{abstract}
Enterprises increasingly seek to query data lakes using natural language via AI-driven tools like semantic operators or deep research agents. However, the latter operates as an opaque black box, hiding its intermediate reasoning and data retrieval steps, and failing to expose controls for managing API costs and execution latency. Meanwhile, the former can be prohibitively expensive for enterprise-scale data lakes. Consequently, analysts using these systems lack the agency to intercept hallucinated premises, verify intermediate results, or correct the system's trajectory. We present \system{}, an interactive execution engine for AI-driven analytics. \system{} compiles natural language requests into physical execution graphs and surfaces them through an interactive notebook interface. Rather than waiting blindly for a final output, users can critique the plan, incrementally execute operators, inspect intermediate data, or directly edit the underlying code or semantic operator instructions. \system{}'s query optimizer will optimize the query with respect to cost or latency constraints provided by the user. Our demo will showcase how \system{} helps users achieve efficient and verifiable insights on workloads motivated by real enterprise use cases.
\end{abstract}

\maketitle

\pagestyle{\vldbpagestyle}
\begingroup\small\noindent\raggedright\textbf{PVLDB Reference Format:}\\
\vldbauthors. \vldbtitle. PVLDB, \vldbvolume(\vldbissue): \vldbpages, \vldbyear.\\
\href{https://doi.org/\vldbdoi}{doi:\vldbdoi}
\endgroup
\begingroup
\renewcommand\thefootnote{}\footnote{\noindent
This work is licensed under the Creative Commons BY-NC-ND 4.0 International License. Visit \url{https://creativecommons.org/licenses/by-nc-nd/4.0/} to view a copy of this license. For any use beyond those covered by this license, obtain permission by emailing \href{mailto:info@vldb.org}{info@vldb.org}. Copyright is held by the owner/author(s). Publication rights licensed to the VLDB Endowment. \\
\raggedright Proceedings of the VLDB Endowment, Vol. \vldbvolume, No. \vldbissue\ %
ISSN 2150-8097. \\
\href{https://doi.org/\vldbdoi}{doi:\vldbdoi} \\
}\addtocounter{footnote}{-1}\endgroup

\ifdefempty{\vldbavailabilityurl}{}{
\vspace{.3cm}
\begingroup\small\noindent\raggedright\textbf{PVLDB Artifact Availability:}\\
The source code, data, and/or other artifacts have been made available at \url{\vldbavailabilityurl}.
\endgroup
}

\begin{figure*}
  \centering
  \includegraphics[width=0.9\linewidth]{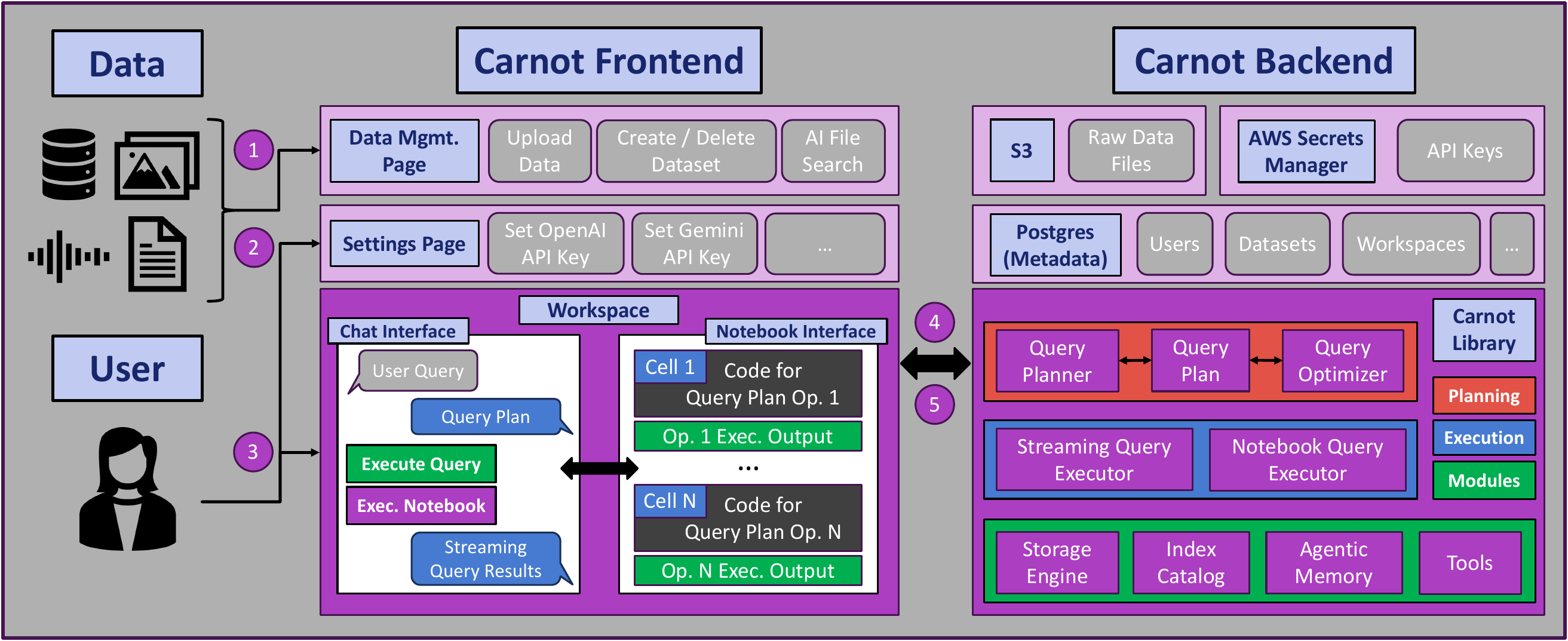}
  \caption{An overview of the \system{} system. Users create datasets from uploaded raw data files (Step 1) and provide API keys which may be used by \system{} to perform LLM-based data processing (Step 2). (Local storage and models are also supported for private deployments.) Users execute queries against their dataset(s) in natural language via a chat interface or via an equivalent representation of the physical plan in a notebook (Step 3). Any feedback provided in the chat and / or edits made to the notebook will be sent to \system{}'s query planner which will re-optimize the query plan in response (Steps 4 and 5).}
  \label{fig:carnot-system}
\end{figure*}

\section{Introduction}
\label{sec:intro}

Organizations increasingly deploy AI-driven query systems to extract insights from unstructured and multi-modal data lakes. For example, a business analyst might issue a natural language query to a Deep Research system which can autonomously explore data, write code, and interpret execution results in a loop to produce a final answer to their question~\cite{openai-deep-research, claude-research, gemini-deep-research, smolagents-open-deep-research, perplexity-deep-research}. The human's job is primarily to verify that the output is correct. However, these systems are often \emph{black boxes}: they accept a query and emit a result, exposing no control over the intermediate chain of retrievals, transformations, and LLM invocations that produced it.

Consider a consumer product analyst who asks an AI system to \emph{``summarize the top complaints about our products and identify which attributes drive negative reviews the most.''}  Suppose the system returns a polished paragraph claiming there are only minor complaints and that three attributes drive most of the negative reviews. The analyst has no way of checking which documents were retrieved, whether the LLM's filter for ``complaints'' silently excluded important data sources, or whether one of the attributes was hallucinated without support from the source data. Any serious analyst, before accepting the result, must carefully read through the reasoning trace (if one is available), and potentially verify parts of the analysis by hand, negating the benefits of automation.

The problem is not merely one of trust; it is one of \emph{control}. In contrast to Deep Research systems, semantic query processing engines~\cite{liu2025palimpzest, russo2025abacuscostbasedoptimizersemantic, shankar2024docetlagenticqueryrewriting, patel2025semanticoperatorsdeclarativemodel, shankar2025docwrangler, wei2026multiobjectiveagenticrewritesunstructured} execute SQL-like semantic query plans. However, once a query is submitted, the user can do little to detect faulty operators mid-execution, apply insights unearthed from partial execution, or selectively patch and re-execute parts of the pipeline. Returning to our example, the analyst would only realize their query is ill-specified \emph{after} paying the full cost of query execution. They must then manually edit their semantic operator code, and discover whether the fix works only after paying a second time.

In short, today's AI-driven query systems fail to provide both \emph{visibility}---the ability to inspect every prompt, code block, and intermediate result before, during, and after execution---and \emph{steerability}---the ability to intervene at any step without discarding prior computation. To address this challenge, we present \system{} (\Cref{fig:carnot-system}), an interactive execution engine for NL queries over unstructured data lakes. When a user issues a natural language query, \system{}'s automated planner compiles it into a Directed Acyclic Graph (DAG) of physical operators, combining semantic operators (e.g., LLM-based filters and extractors) with agentic operators (e.g., data exploration, Python code generation and execution). Unlike prior semantic operator systems, \system{} materializes the DAG as a notebook interface where each cell corresponds to a single operator. Users can inspect any cell at multiple levels of abstraction---a summary of the operator's intent, the generated code or prompt, or the raw input/output data pairs---and execute the graph one operator at a time. When a user edits a cell to correct a mistake, \system{}'s optimizer will automatically re-optimize and re-execute the cell while reusing the results of unaffected upstream cells.

This feedback loop is the core technical enabler of \system{}'s interactive model. In our previous example, the analyst first inspects the proposed plan of the agent and corrects any faulty assumptions in plan-space. They then execute the first few operators and discover that the LLM's filter only looks at the ``ratings'' column and excludes the free-form ``review text'' field. They edit the filter prompt directly in its notebook cell. \system{} resumes execution from the corrected point---leaving the upstream data retrieval and preprocessing steps untouched. The result is a system that combines the optimizable structure of a database query plan with the iterative flexibility of an agent, while keeping the human in the loop at every step.

Our demo showcases \system{}'s interactive planning and execution over datasets from academic benchmarks~\cite{lai2026kramabench} and enterprise workloads motivated by industry partners. To see transparent execution in action, attendees will issue queries and inspect the resulting DAGs before any computation begins; they may drill into any operator to view its generated code, prompt, or a natural language summary of its intent. To experience steerability, attendees will execute a query step-by-step, intercept a faulty intermediate result, and edit the corresponding notebook cell; \system{} will then resume execution from the corrected point without re-running upstream operators. Finally, to highlight \system{}'s cost-aware optimization, attendees will configure explicit cost and latency constraints in the interface; \system{} will show how the planner dynamically swaps operator implementations to meet their budget.

\section{System Overview}


\subsection{\system{} Frontend}
\label{sec:frontend}

\system{}'s data management page enables users to upload raw data files (e.g., a \texttt{.zip} archive) to \system{}'s data management page and organize them into one or more named \emph{datasets}---for example, a collection of research papers, a set of supplier invoices, or a dataset of product reviews (Step~1 in \cref{fig:carnot-system}). Each dataset may be annotated with a description of its contents, which \system{}'s query planner consults when generating execution plans. The user also registers API keys for one or more supported LLM providers (Step~2).

To issue a query, the user selects the relevant datasets and types a natural language question into the chat interface (Step~3). \system{}'s backend responds with a physical query plan, displayed in complementary views: a linearized bullet-list summary in the chat window, and a visual DAG beside it. Crucially, execution has not yet begun---\system{} pauses here to let the user critique the proposed plan before any additional tokens are spent. The user may issue follow-up instructions to revise the plan, execute it end-to-end with results streamed back to the chat, or open it as an interactive \emph{notebook}.

The notebook is the primary vehicle for \system{}'s visibility and steerability guarantees. Each cell corresponds to a single physical operator in the DAG, obtained by topologically sorting the plan so that data-source operators appear first. Users can inspect any cell at three levels of abstraction: a natural language summary of the operator's intent, the generated code or LLM prompt, and the raw input/output data pairs. Operators can be executed one cell at a time, and any cell may be edited in-place---for instance, to revise a semantic filter's predicate or to swap an LLM call for a deterministic Python expression. When the user commits an edit, \system{} re-executes the updated operator while reusing the cached results of unaffected upstream cells (Steps~4 and~5). Users may also set explicit cost and latency budgets; \system{}'s optimizer respects these constraints by dynamically selecting cheaper or faster operator implementations.

\subsection{\system{} Backend}
\label{sec:backend}

The backend is responsible for managing the metadata and state that support the frontend and dispatching query planning, optimization, and execution requests to the \system{} library. For the former, \system{} stores user identifiers, dataset definitions, and the state of each conversation--notebook pair (together called a \emph{workspace}) in a Postgres database. User-uploaded data files are placed in a private S3 bucket. (\system{}'s storage and LLM abstractions can be swapped for local filesystem(s) and local models in deployments where privacy is a concern.) Using stored state, the backend translates user actions into calls against the \system{} library, described next. 

\subsection{\system{} Library}
\label{sec:library}

The library provides modular components for planning, optimizing, and executing queries. Its core abstraction is a \texttt{Dataset} class---a wrapper around a list of files that exposes a set of chainable operators. These include \emph{semantic operators} (LLM-based maps, filters, joins, etc.) and \emph{agentic operators} (data exploration, Python code generation and execution, general-purpose reasoning). A user or agent constructs a query plan by chaining operators into a DAG over one or more input datasets; data is only materialized when \texttt{Dataset.run()} is called. The \texttt{Dataset} class is JSON serializable, which enables it to be directly visualized and edited in the notebook.

The \texttt{QueryPlanner} is an LLM agent with few-shot demonstrations for composing plans in \system{}'s operator vocabulary. To avoid premature use of expensive LLM operators, the planner first invokes a \emph{data discovery} subagent that samples the dataset, performs schema analysis, and summarizes dataset structure. The \texttt{QueryOptimizer} follows, inspired by the Cascades-style optimizer of Abacus~\cite{russo2025abacuscostbasedoptimizersemantic}. It shares a similar set of implementation and transformation rules, but differs in its use of an LLM agent to select the final physical plan from the estimated Pareto frontier of cost, latency, and quality. The library also contains a streaming and notebook execution layer, a storage engine with caching support, and an index catalog for managing semantic indices.

\subsection{Putting It Together}
\label{sec:walkthrough}

\begin{figure*}
  \centering
  \includegraphics[width=\linewidth]{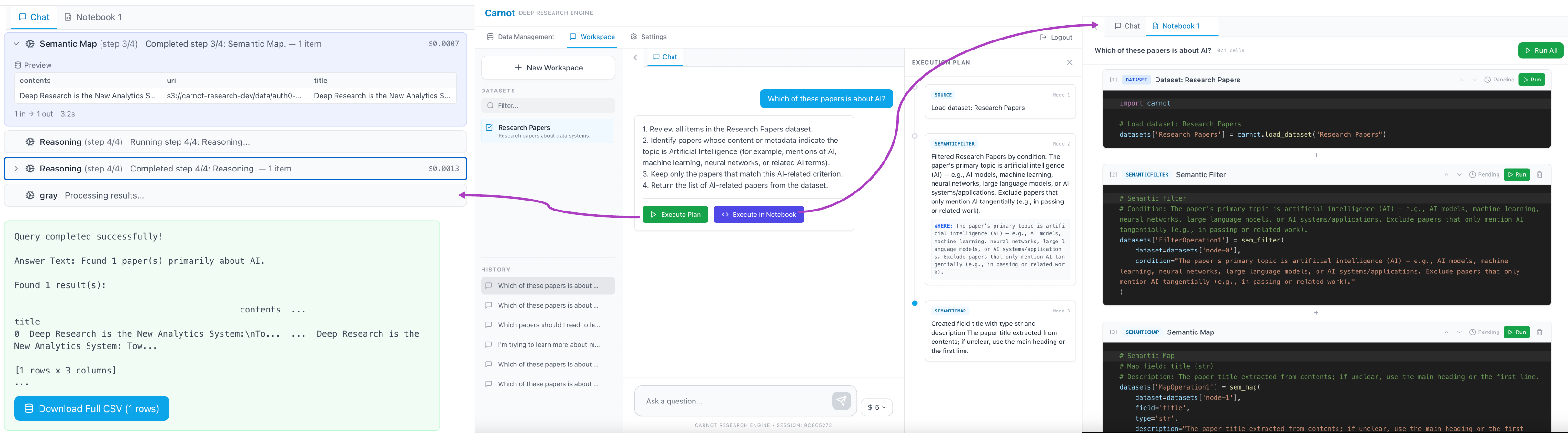}
  \caption{A \system{} workspace answering the query \emph{``Which of these papers is about AI?''} Center: the chat interface with the linearized plan and visual DAG. Right: the notebook's code cells, exposing the \system{} library calls that implement each operator. Left: the finished query execution showing per-operator results, cost annotations, and the final output.}
  \label{fig:carnot-demo}
\end{figure*}

\Cref{fig:carnot-demo} illustrates a complete \system{} session for the query \emph{``Which of these papers is about AI?''} In the center panel, the chat interface displays the query alongside the linearized plan and visual DAG; the user can inspect the plan's logic before choosing to execute it in the chat or open it as a notebook. On the right, the notebook's code cells reveal the \system{} library calls---including the semantic filter's condition and the semantic map's field description---giving the user full transparency into each operator's behavior. On the left, the chat interface shows per-operator results with cost annotations (e.g., \$0.0087 for the semantic map step), a final answer, and a downloadable CSV of the output (similar outputs are available in the notebook). 

\section{Demonstration Scenarios}
\label{sec:demo}
Our demonstration is organized around three scenarios that let attendees experience Carnot's capabilities first-hand on the provided datasets, using the interface shown in \Cref{fig:carnot-demo}.

\subsection{Datasets}
We will provide attendees with access to two datasets. The first dataset is an anonymized set of 1,000 consumer product reviews from an industry partner. We anonymized the reviews by stripping the real product names and product identifiers from the dataset, while also having an LLM paraphrase the review text and titles. We inspected the anonymized reviews to ensure that their sentiment (and content) remained faithful to their original form. The second dataset will be the \texttt{legal} workload from KramaBench~\cite{lai2026kramabench} an academic benchmark for data analytics.

 
\subsection{Transparent Plan Inspection}
\label{sec:demo-inspect}
 
This scenario highlights \system{}'s ability to inspect every operator before any computation begins. The attendee will issue a free-form natural language query---for example, \emph{``summarize the top complaints about our products and identify which attributes drive negative reviews the most.''} against our enterprise dataset. \system{} will return a logical plan displayed both as a nested summary in the chat and as a DAG. The attendee will examine the DAG of operators, including per-operator arguments such as filter conditions and group by fields, in order to understand the logic of the query plan. Before executing a single operator, the attendee can verify that the plan retrieves data from the intended sources, that the filter criteria match their expectations, and that the aggregation logic is sound.

 
\subsection{Steering via the Chat Interface}
\label{sec:demo-chat}
 
This scenario demonstrates high-level \emph{steerability} through
conversational feedback. Continuing from the plan above, the attendee notices the filter operator only inspects the ``ratings'' field.  Rather than editing the operator directly, they type a follow-up instruction in the chat: \emph{``Be sure to also consider any negative comments left in the `review text' field.''} \system{}'s planner treats this instruction as a constraint, re-plans the affected operators, and presents an updated DAG in the chat. The attendee can then execute the revised plan in the chat, streaming per-operator results back to the chat interface. If the results look satisfactory, they accept the output. If further refinement is needed, they may continue the conversation---or drill down into different views for fine-grained control. 

\subsection{Editing via the Notebook Interface}
\label{sec:demo-notebook}
 
This scenario showcases \system{}'s most distinctive capability: cell-level editing with query re-execution and re-optimization. The attendee opens the current query plan as a notebook, executes the first few cells, and inspects the intermediate outputs. Suppose the semantic filter's LLM prompt produces false negatives---it classifies legitimate complaints as neutral feedback. The attendee edits the cell directly to make sure the prompt instruction captures forms of feedback it previously misclassified. Upon saving the edit, the updated cell appears in the notebook and \system{}'s optimizer re-optimizes the implementation of the corresponding operator. Upstream cells whose results remain valid are \emph{not} re-executed.  The attendee then resumes execution from the corrected cell, iterating on individual operators until the final output meets their standards.

To highlight cost-aware optimization, the attendee will also configure explicit cost and latency constraints via the interface. \system{} will show how the planner dynamically swaps operator implementations---e.g., replacing a large frontier model with a smaller, cheaper one for a low-precision filter---to meet the specified budget while preserving result quality where it matters most.


\section{Limitations and Future Work}
\system{} does not provide guarantees on plan accuracy or correctness. We could modify the optimizer to provide statistical guarantees with respect to an oracle \cite{patel2025semanticoperatorsdeclarativemodel}, but this is a soft guarantee since the oracle can also be wrong. Alternatively, better interfaces may improve users' ability to verify plan correctness. In the future, we also plan to extend \system{} to support search over the web (akin to traditional Deep Research) and general tool calling. We also aim to explore improvements to the notebook interface which can better highlight the non-linear nature of most plan DAGs.

\begin{acks}
We are grateful for the support from the DARPA ASKEM Award HR00112220042, the ARPA-H Biomedical Data Fabric project, NSF DBI 2327954, a grant from Liberty Mutual, a Google Research Award, and the Amazon Research Award. Additionally, our work has been supported by contributions from Amazon, Google, and Intel as part of the MIT Data Systems and AI Lab (DSAIL) at MIT, along with NSF IIS 1900933. This research was sponsored by the United States Air Force Research Laboratory and the Department of the Air Force Artificial Intelligence Accelerator and was accomplished under Cooperative Agreement Number FA8750-19-2-1000. The views and conclusions contained in this document are those of the authors and should not be interpreted as representing the official policies, either expressed or implied, of the Department of the Air Force or the U.S. Government. The U.S. Government is authorized to reproduce and distribute reprints for Government purposes notwithstanding any copyright notation herein.
\end{acks}


\bibliographystyle{ACM-Reference-Format}
\bibliography{sample}

\end{document}